# Peculiarities of polarization switching in ferroelectric semiconductors with charged inhomogeneities.

Anna N. Morozovska[1], Eugeny A. Eliseev[2]

[1]V.Lashkaryov Institute of Semiconductor Physics, NAS of Ukraine,
41, pr. Nauki, 03028 Kiev, Ukraine,
e-mail: morozo@mail.i.com.ua

[2]Institute for Problems of Materials Science, National Academy of Science of Ukraine,
Krjijanovskogo 3, 03142 Kiev, Ukraine,

We have proposed the phenomenological description of polarization switching peculiarities in ferroelectric semiconductors with charged defects and prevailing extrinsic conductivity. Exactly we have modified Landau-Ginsburg approach shown that the macroscopic state of the aforementioned inhomogeneous system can be described by three coupled equations for three order parameters. Both the experimentally observed coercive field values well below the thermodynamic one and the various hysteresis loop deformations (minor, constricted and double loops) have been obtained in the framework of our model. The obtained results qualitatively explain the ferroelectric switching in such bulk ferroelectric materials as SBN single crystals doped with Ce, PZT films doped with Nd and La- doped PZT ceramics.

## 1. Introduction

The spontaneous electric displacement switching under the external field is one of the key feature of the ferroelectric materials. The ferroelectric hysteresis loop (or the ambiguous dependence of the electric displacement on the external electric field) is widely used in applications [1]. The main comprehensively studied characteristics of hysteresis loop are the spontaneous displacement and coercive field values. However there is a discrepancy between experimental and theoretical results. Even for the perfect bulk ferroelectric materials the classical theoretical models give only the qualitative picture of the displacement switching, but quantitative characteristics, namely the theoretical value of coercive field is essentially greater than the experimentally observed values. As for the imperfect ferroelectrics, here the hysteresis loop looks rather slim in comparison with the theoretical square one, and its shape undergoes various deformations.

Unsaturated hysteresis loops are widely observed both in ferroelectrics doped with such rare-earth metals as La, Ce, Nd and in PZT films with such unavoidable technological defects as oxygen or lead vacancies (see e.g. [2], [3], [4], [5]). The change of the saturation mechanism may be related to the aforementioned dopants influence or imperfections appearance, owing to the facts that its characteristic strongly depends on the dopant concentration [3] or deposition temperature [4].

Sometimes the minor hysteresis loops appear in doped ferroelectrics (see e.g. [6], [7], [8]). This phenomena is strongly dependent on the type and concentration of dopants [7] and the applied field frequency [6], [7].

Another interesting feature of the polarization switching in ferroelectrics with non-isovalent dopants is the constricted hysteresis loops observed in Nd- doped [3] and La-doped lead zirconate titanate [9]. Constricted hysteresis loops similar to observed by authors of ref. [9] were found in the La-doped lead zirconate titanate ceramics with composition near the morphotropic boundaries between ferroelectric tetragonal, rhombohedral and antiferroelectric phases [10].

It is generally accepted that the ferroelectrics inhomogeneities promote the domains nucleation, because the estimation was shown that the energetic barrier is too high for the nucleation in the homogeneous system [11]. On the other hand it was shown recently, that for the ferroelectric film with "dead" layers appearance of several nuclei do not need to overcome the energetic barrier due to the long range interaction between nuclei [12].

It should be noted, that although using the Kolmogov-Avraami theory to model ferroelectric displacement reversal allows one to describe experimental data with high accuracy (see e.g. [13]), one has to use fractional nuclei dimension [14]. Furthermore, the nuclei long-range interactions are neglected in models of Kolmogov-Avraami type, and question about nuclei appearance is not discussed [12].

The phenomenological Landau-Ginsburg-Devonshire theory evolved for the mono-domain perfect ferroelectrics predicts that the homogeneous displacement switching takes place when the external field reaches so called thermodynamic coercive field (see e.g. [15]). At this point system with displacement vector opposite to the external field looses stability, so with the external field absolute value increase homogeneous switching of displacement takes place. However, the thermodynamic coercive field, which is proportional to product of the spontaneous displacement and susceptibility, usually is from several times to several orders of magnitude greater than experimental value for the bulk systems. In real system domains with opposite polarization (nucleus) appears at much smaller external field values than thermodynamic coercive field [15].

One of the obvious way to solve the problem is to take into account the correlation or gradient energy in the free energy of Landau-Ginsburg-Devonshire (see e.g. [16], [17]). This additional term directly corresponds to the surface energy of the domain walls. However, in the homogeneous media appearance of the domains walls leads to the free energy increase [16], [17]. Therefore this model describes the switching without domains (so called homogeneous switching) and neither domain pinning, nor domain nucleation or domain movement and do not allow to describe inhomogeneous switching in bulk material. For the finite size system (e.g. film) the model based on the inhomogeneous Landau-Ginsburg-Devonshire free energy predicts that coercive field increase or decrease with size increase depending on the boundary conditions (see e.g numerical calculation in [18]). This result can be explained with the help of the effective free energy depending on the film thickness which was derived recently [19, 20] for monodomain ferroelectric film allowing for the depolarization field. Nevertheless, it should be once more stressed that even in this model the coercive field tends to the thermodynamic coercive field with film thickness increase. However, the coercive field value experimentally observed for the ferroelectric films sometimes considerably increases with film thickness decrease (see e.g. [21], [22]).

In spite of this discrepancy, i.e. inapplicability of the classical Landau-Khalatnikov equation to the processes of spontaneous induction switching, the question of the inhomogeneous ferroelectrics switching and the modification of Landau-Khalatnikov equation to the description of these processes is remained unsolved. To our mind, first of all the random electric field arising near charged inhomogeneities should be taken into account. However in contrast to the random field theory, developed for the relaxor insulator ferroelectrics [23], semiconductor properties of the material (at least extrinsic conductivity created by charged impurities) should be taken into consideration. In this case inhomogeneous electric field should be obtained self consistent from electrostatic Maxwell equations allowing for the random fluctuations of free carriers density near defects, electric current continuity equation and equation of state for the polar nonlinear media. Right in the way we modified Landau-Ginsburg approach for the ferroelectrics-semiconductors with charged defects and found both the essential coercive field decrease and the hysteresis loops deformation experimentally observed in inhomogeneous systems.

The main goal of the paper is to demonstrate that macroscopic state of the bulk sample with random charged defects can be described by the system of three coupled equations, which is similar to such well-known nonlinear systems of first order differential equations as the Lorenz one [24]. Such dynamical systems of equations could reveal chaotic regimes, strange attractors as well as strongly non-ergodic behaviour and continuous relaxation time spectrum. Until now, we have simulated only the quasi-equilibrium ferroelectric hysteresis loops. Certainly, the dynamical dielectric response of this first obtained system requires further investigations.

## 2. The problem

We would like to underline that the majority of doped ferroelectrics posses conductivity at room temperature. In all aforementioned materials where footprint, unsaturated loops and minor loops exist, either non-isovalent impurities or some unavoidable imperfections manifest themselves as charged defects. Moreover, these "dusty" materials would rather be considered as extrinsic semiconductors [25, 26] than perfect insulators. These two facts give the basis of proposed phenomenological model.

We assume that almost immovable non-stechiometric defects or non-isovalent (photoactive) impurity centers are embedded into hypothetical perfect uniaxial ferroelectric (z direction coincides with the polar axis one). We suppose that impurity centers or defects are ionized (e.g. after UV, photo- or thermal excitation) and even in the absence of proper conductivity, they provide a prevailing extrinsic (or photo-) conductivity in the bulk sample.

Thus we regard that the sample as a whole is the electro-neutral extrinsic semiconductor with prevailing $n$-type conductivity and positively charged defects with density $\rho_s(\mathbf{r})$. The microscopic spatial distribution of these defects charge density $\rho_s(\mathbf{r})$ is characterized by the average charge density $\overline{\rho}_s$ proportional to the ionized defects concentration and microscopic modulation $\delta\rho_s$, i.e. $\rho_s(\mathbf{r}) = \overline{\rho}_s + \delta\rho_s(\mathbf{r})$. Hereinafter $d$ is the average distance between defects, $2r_0$ is the average size of defects (see Scheme 1).

The modulation $\delta\rho_s$ is not regular periodic function with spatial period $d$ but randomly fluctuates due to the great variety of misfit effects (lattice distortions, local shift, possible clusterization at high defect concentration).

The movable screening clouds $\delta n(\mathbf{r},t)$ surround each charged center (see Scheme 1a). The characteristic size of these screening clouds is of the same order as the Debye screening radius $R_D$. When one applies the external field $E_0$, screening clouds of free charges are deformed, and nano-system "defect center + screening cloud" becomes polarized (Scheme 1b). Polarized regions "$\delta\rho_s(\mathbf{r})+\delta n(\mathbf{r},t)$" cause the additional inner electric field fluctuations $\delta E(\mathbf{r},t)$. According to the equations of state, the fluctuations of the inner electric field $\delta E$ cause displacement fluctuations $\delta D$. The modulation $\delta\rho_s$ is not regular periodic function with spatial period $d$ but randomly fluctuates due to the great variety of misfit effects (lattice distortions, local shift, possible clusterization at high defect concentration).

The movable screening clouds $\delta n(\mathbf{r},t)$ surround each charged center (see Scheme 1a). The characteristic size of these screening clouds is of the same order as the Debye screening radius $R_D$. When one applies the external field $E_0$, screening clouds of free charges are deformed, and nano-system "defect center + screening cloud" becomes polarized (Scheme 1b). Polarized regions "$\delta\rho_s(\mathbf{r})+\delta n(\mathbf{r},t)$" cause the additional inner electric field fluctuations $\delta E(\mathbf{r},t)$. According to the equations of state, the fluctuations of the inner electric field $\delta E$ cause displacement fluctuations $\delta D$.

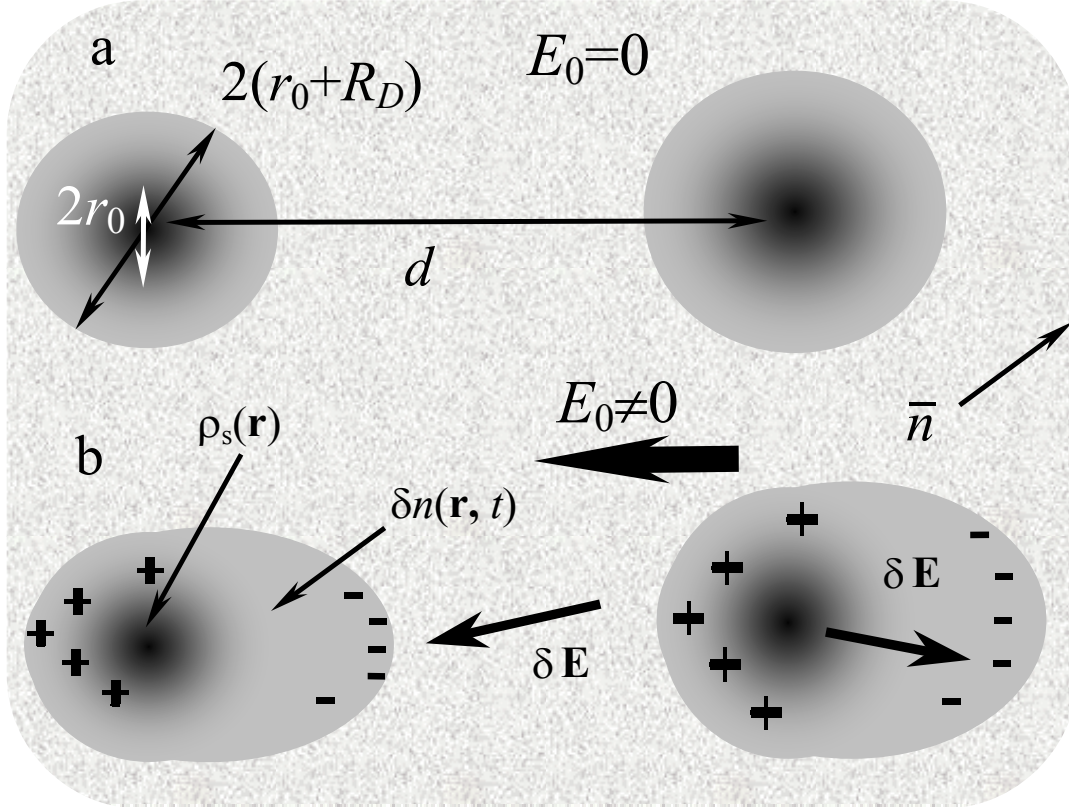

**Scheme 1.** The charged defects with the charge density $\rho_s$ (dark circles with radius $r_0$) are screened by the free charges with density $\delta n$ (gray circles or ellipses with screening radius $R_D$). The parts "a" and "b" show the system with the zero and nonzero external field $E_0$ respectively.

Evidently non-homogeneous mechanical stresses and local symmetry distortion appeared near the defects should be taken into account [16], [17]. The consideration of these effects significantly complicates the problem, and we hope that the system behavior will not change qualitatively under their influence. Also it is known (see e.g. [27]) that homogeneous elastic stresses due to the electrostriction coupling with the polar order parameter can be taken into account by the renormalization of the free energy expansion coefficients.

### 3. General equations

Maxwell's equations for the electric displacement **D**, field **E** and equation of continuity have the form:

$$div\,\mathbf{D} = 4\pi\rho, \quad rot\,\mathbf{E} = 0, \quad \frac{\partial \rho}{\partial t} + div\,\mathbf{j}_c = 0 . \quad (1)$$

They have to be supplemented by the equations of state:

$$\mathbf{j}_c = \sum_m \left(\mu_m \rho_m \mathbf{E} - \kappa_m grad\,\rho_m\right), \quad \rho(\mathbf{r},t) = \sum_m \rho_m(\mathbf{r},t) + \rho_s(\mathbf{r}). \quad (2)$$

Here $\rho_m$, $\mu_m$ and $\kappa_m$ are the $m$-type movable charge volume density ($m=n,p$), mobility and diffusion coefficient respectively, $\mathbf{j}_c$ is the macroscopic free-carriers current, $\rho_s(\mathbf{r})$ is the fluctuating charge density of static defects.

Keeping in mind that the sample is the extrinsic semiconductor with prevailing *n*-type conductivity, hereinafter we neglect the proper conductivity, put $n \approx \sum_m \rho_m$ and omit the subscript "*m*". So equations (1), (2) can be rewritten as:

$$div\mathbf{D} = 4\pi(n + \rho_s),$$

$$div\left[\mu\, n\, \mathbf{E} - \kappa\, grad\, n + \frac{1}{4\pi}\frac{\partial}{\partial t}\mathbf{D}\right] = 0. \tag{3}$$

Here $n<0$, $\mu<0$ and $\kappa>0$. In accordance with Einstein relation $\mu/\kappa \approx e/k_B T^*$ ($e<0$) the Debye screening radius $R_D = \sqrt{\kappa/4\pi n\mu} \approx \sqrt{\left|k_B T/4\pi ne\right|}$. Hereafter we suppose that homogeneous external field $E_0(t)$ is applied along polar *z* - axis. The sample occupies the region $-\ell < z < \ell$, i.e. it is infinite in the transverse directions. Let us consider that the electrodes potential difference $\varphi = 2\ell E_0(t)$ at $z = \pm\ell$ which is independent on transverse coordinates. Therefore, the inner field satisfies the conditions:

$$\frac{1}{2\ell}\int_{-\ell}^{\ell} E_z(\mathbf{r},t)\,dz = E_0(t). \tag{4}$$

Boundary conditions depend on the mechanism of the spontaneous displacement screening [28], associates with the formation of oppositely charged surface charged layers with thickness of several lattice constants. We assume that the displacement distribution is symmetrical for a rather thick sample with equivalent boundaries $z = \pm\ell$. Also we introduce the averaging over sample volume:

$$f(\mathbf{r},t) = \overline{f(t)} + \delta f(\mathbf{r},t), \quad \overline{f(t)} = \frac{1}{V}\int_V f(\mathbf{r},t)d\mathbf{r}. \tag{5}$$

Hereinafter the dash designates the averaging over sample volume $V$, $f = \{n, \rho_s, E, D, j, ...\}$, by definition $\overline{\delta f(\mathbf{r},t)} = 0$. The spatial distribution of the deviations $\delta f(\mathbf{r},t)$ consist of the part caused by spontaneous displacement screening and localized in the ultrathin screening surface regions [28] and the fluctuations caused by microscopic modulation $\delta\rho_s$. For the μm-thick sample the contribution from the ultrathin screening region to the average functions $\overline{\delta f^m(\mathbf{r},t)}$ is negligibly small, and one can regard that:

$$\overline{\delta f^{2m+1}(\mathbf{r},t)} \approx 0, \qquad m = 1; 2... \tag{6}$$

Also we suppose that the correlation between the different $\delta f$-functions is equal to zero if the total power of the functions is an odd number. It follows from (3)-(5) that:

$$\mathbf{D}(\mathbf{r},t) = \mathbf{e}_z\overline{D(t)} + \delta\mathbf{D}(\mathbf{r},t), \tag{7a}$$

$$\mathbf{E}(\mathbf{r},t) = \mathbf{e}_z E_0(t) + \delta\mathbf{E}(\mathbf{r},t). \tag{7b}$$

Here $\mathbf{e}_z$ s the unit vector directed along z-axis. $\overline{E}$ is the applied uniform field $E_0(t)$ and $\overline{E}_{x,y} = 0$. Notice that the average values $\overline{E}, \overline{D}$ are determined experimentally [28], [29] most of the times. Having substituted (7) into (3) and averaged, one can obtain the expressions for the average quantities, namely:

$$\overline{\rho} = 0 \quad \Rightarrow \quad \overline{n} = -\overline{\rho}_s, \quad \overline{\mathbf{j}}(t) = \overline{\mathbf{j}}_c(t) + \mathbf{e}_z\frac{\partial}{\partial t}\frac{\overline{D(t)}}{4\pi}. \tag{8}$$

The absence of the space charge average density $\overline{\rho}$ follows from the sample electro-neutrality and corresponds to the result [28], [30]. Here $\overline{\mathbf{j}}(t)$ is the total macroscopic current. Using (5), (7) one can obtain that

$$div(\delta \mathbf{D}) = 4\pi(\delta n + \delta \rho_s(\mathbf{r})), \tag{9}$$

$$div\left(\mu\left[\delta n\, E_0 \mathbf{e}_z + \left(-\overline{\rho}_s + \delta n\right)\delta \mathbf{E}\right] - \kappa\, grad\, \delta n + \frac{1}{4\pi}\frac{\partial}{\partial t}\delta \mathbf{D}\right) = 0. \tag{10}$$

As the equation of state, we use the Landau-Khalatnikov equation for the displacement z-component relaxation, but take into account the influence of random electric field $\delta E_z$ created by charged defects and correlation effects. This approach takes into consideration the spatial-temporal dispersion of the ferroelectric material. Using the original approach evolved in paper [31] and formula (7), we modify classical Landau-Khalatnikov equation $\Gamma \frac{\partial}{\partial t} D_z + \alpha D_z + \beta D_z^3 = E_z$ and obtain the following system of coupled equations:

$$\Gamma \frac{\partial \overline{D}}{\partial t} + \left(\alpha + 3\beta \overline{\delta D^2}\right)\overline{D} + \beta \overline{D}^3 = E_0(t), \tag{11}$$

$$\Gamma \frac{\partial}{\partial t}\delta D + \left(\alpha + 3\beta \overline{D}^2\right)\delta D + 3\beta \overline{D}\left(\delta D^2 - \overline{\delta D^2}\right) + \beta \delta D^3 - \gamma \frac{\partial^2 \delta D}{\partial \mathbf{r}^2} = \delta E_z. \tag{12}$$

Here $\Gamma > 0$ is the kinetic coefficient, $\alpha < 0$, $\beta > 0$, $\gamma > 0$ are parameters of the hypothetical pure (free of defects) sample. Hereafter we denote $\delta D_z \equiv \delta D$. We would like to underline, that the sum of equations (11)-(12) coincides with Landau-Khalatnikov equation [15], [18] for $D$ only under the condition $\delta \rho_s = 0$. This condition corresponds to the absence of random inhomogeneities or their seeding, and thus only the homogeneous polarization switching can take place when the external field exceeds thermodynamic coercive field. In such case the inner field in the bulk of the sample is the sum of the external field and predetermined depolarization field originated from displacement screening [30]. In contrast to this at $\delta \rho_s \neq 0$ the inner field $\delta \mathbf{E}$ contains the random component dependent over $\delta \rho_s(\mathbf{r})$ and $\delta D$ in accordance with (9)-(10).

The system of equations (9)-(12) is complete, because the quantities $\delta n, \delta \mathbf{E}$ can be expressed via the fluctuations of displacement $\delta D$ and $\delta \rho_s(\mathbf{r})$ allowing for (9), (10). System (9)-(12) determines the spatial-temporal evolution of the displacement in the bulk sample and has to be supplemented by the initial distributions of all variables.

Hereinafter we consider only the average characteristics of the ferroelectric semiconductors with non-isovalent impurities. The study of the mechanisms of domain wall pinning by the given distribution of charged defects, domain nucleation during spontaneous displacement reversal is beyond the scope of this paper. Such problems for the ferroelectrics - ideal insulators were considered in details earlier (see e.g. [11], [32]).

## 4. Coupled equations.

In order to simplify the nonlinear system (9)-(12) the following assumptions have been used.

1) The charged inhomogeneities are surrounded by screening clouds, therefore $\delta \rho_s \sim \delta n$ and

$$\overline{\delta \rho_s \delta n} \approx -\eta \overline{\delta \rho_s^2}, \qquad 0 < \eta < 1. \tag{13}$$

At the small amplitude of external field the positive function $\eta$ is determined by the ratio $R_D/r_0$ (see Fig. 1 and Appendix A).

When external field amplitude increases, $\eta$ value slightly decreases due to the polarization of the system "charged fluctuation + screening cloud". We suppose that $\eta$ can be approximated by effective constant value at not very high external field. Note, that typical defect concentrations ~1-10% provide high enough extrinsic conductivity at room temperature, and thus the average distance $d$ between inhomogeneities is greater then the Debye screening radius $R_D$, namely: $R_D \sim 1$ nm [28], [30], $d \sim 5$nm (see Fig. 1).

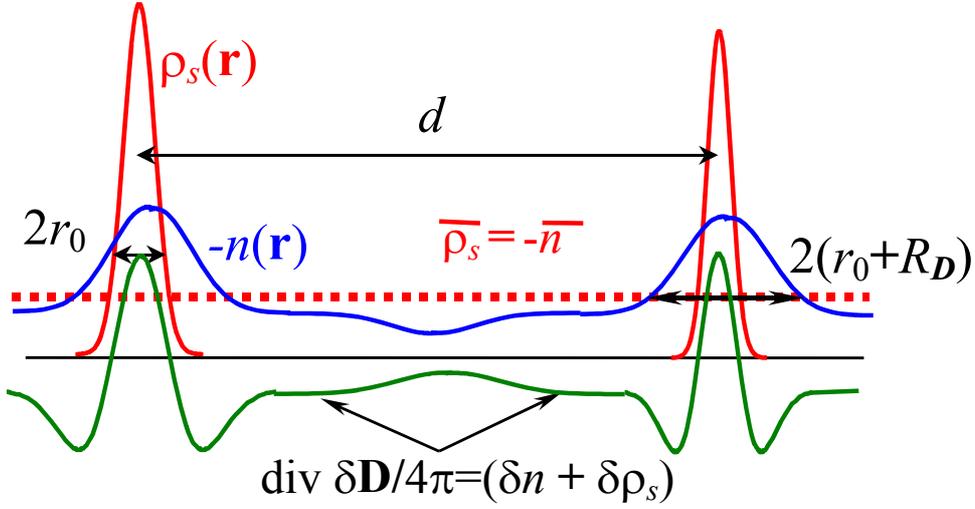

**Figure 1.** The screening of the charged defects $\delta \rho_s$ by free charges $\delta n$ at small amplitude of external field.

2) The fluctuating microscopic circular currents $rot\,\delta \psi$ around charged defects could be neglected in comparison with other currents $\delta \mathbf{j}$:

$$\left| \delta \mathbf{j} \right| >> \left| rot\,\delta \psi \right|. \tag{14}$$

After elementary transformations of (10) allowing for (14) the equation for electric field fluctuation $\delta \mathbf{E}$ acquires the form:

$$\delta \mathbf{E} \approx \frac{1}{4\pi\mu\,\overline{\rho}_s} \frac{\partial}{\partial t} \delta \mathbf{D} + \mathbf{e}_z \frac{\delta n}{\overline{\rho}_s} E_0(t) - \frac{\kappa}{\mu\overline{\rho}_s} grad\,\delta n + \frac{\delta n\,\delta \mathbf{E} - \overline{\delta n\,\delta \mathbf{E}}}{\overline{\rho}_s}. \tag{15}$$

The term $\overline{\delta n\,\delta \mathbf{E}}$ does not contribute into (10), but provides the condition $\overline{\delta \mathbf{E}} = 0$. The equation (9) gives

$$\delta n = \frac{1}{4\pi} div(\delta \mathbf{D}) - \delta \rho_s(\mathbf{r}). \tag{16}$$

Using (16) and (15) the field electric field fluctuations $\delta \mathbf{E}$ caused by charged defects can be expressed via $\delta D$ and $\delta \rho_s$:

$$\delta \mathbf{E} \approx \frac{\partial}{\partial t} \frac{\delta \mathbf{D}}{4\pi\mu\overline{\rho}_s} + \left( \mathbf{e}_z E_0(t) - \frac{\kappa\,grad}{\mu} \right) \left( \frac{div(\delta \mathbf{D})}{4\pi\overline{\rho}_s} - \frac{\delta \rho_s}{\overline{\rho}_s} \right) + $$
$$+ \frac{\overline{\delta \mathbf{E}\,div(\delta \mathbf{D})} - \delta \mathbf{E}\,div(\delta \mathbf{D})}{4\pi\overline{\rho}_s} + \frac{\delta \mathbf{E}\,\delta \rho_s - \overline{\delta \mathbf{E}\,\delta \rho_s}}{\overline{\rho}_s} \tag{17}$$

The equations (11), (12), and (17) is the self-consistent system of the nonlinear integral-differential equations for $\delta D$, $\delta E$ and $\overline{D}$. Its non-homogeneity is proportional to charge fluctuations $\delta \rho_s$ and external field $E_0$.

The approximate system of first-order differential equations for average displacement $\overline{D}$, its mean-square fluctuation $\overline{\delta D^2}$ and correlation $\overline{\delta D\,\delta \rho_s}$ can be derived after some elementary transformations of (11), (12) (see Appendix B and (13), (17)). Thus, we obtain three coupled equations:

$$\Gamma\frac{\partial\overline{D}}{\partial t}+\left(\alpha+3\beta\overline{\delta D^2}\right)\overline{D}+\beta\overline{D}^3=E_0(t),\tag{18a}$$

$$\frac{\Gamma_R}{2}\frac{\partial}{\partial t}\overline{\delta D^2}+\left(\alpha_R+3\beta\overline{D}^2\right)\overline{\delta D^2}+\beta\left(\overline{\delta D^2}\right)^2=-E_0(t)\frac{\overline{(\delta\rho_s\delta D)}}{\overline{\rho}_s}-\left(4\pi R_D\right)^2(1-\eta)\overline{\delta\rho_s^2},\tag{18b}$$

$$\Gamma_R\frac{\partial}{\partial t}\overline{\delta D\delta\rho_s}+\left(\alpha_R+3\beta\overline{D}^2+\beta\overline{\delta D^2}\right)\overline{\delta D\delta\rho_s}=-E_0(t)\frac{\overline{\delta\rho_s^2}}{\overline{\rho}_s}\eta.\tag{18c}$$

Here the renormalized coefficients $\alpha_R=\alpha+\left(\gamma+R_D^2\right)/d^2$ and $\Gamma_R=\Gamma-1/4\pi\mu\overline{\rho}_s$ have been introduced. The renormalization of $\alpha$ takes into account the finite value of correlation length $\ell_c=\sqrt{\gamma/|\alpha|}$ and classical renormalization of gradient term $\gamma$ by Debye screening radius $R_D$ in the bulk of a sample [28], [30]. Moreover, for the typical values of parameters the renormalized coefficient $\alpha_R$ is positive (e.g. in CGSE system $\alpha\sim$ -2·10$^{-3}$, $\gamma\sim$10$^{-15}$ cm$^2$ [15, 17, 29], defects concentration ~1% provides $d\sim$5·10$^{-7}$ cm and so $\alpha_R\sim$ +2·10$^{-3}$). The renormalization of $\Gamma$ takes into account the finite value of maxwellian relaxation time $\tau_m=-1/4\pi\mu\overline{\rho}_s$. The additional source of displacement fluctuations is the term $-4\pi(1-\eta)\kappa\overline{\delta\rho_s^2}/\overline{\rho}_s\mu\approx16\pi^2R_D^2(1-\eta)\overline{\delta\rho_s^2}$ in the right-hand side of (18b). It is proportional to defects concentration and originated from diffusion field $\kappa\,grad(n)/\mu\rho_s$ (see (10) and (15)). All these effects have been neglected in the classical equations of Landau-Khalatnikov type as well as in our previous papers [33]-[34].

The system (18) determines the temporal evolution of the bulk sample dielectric response and have to be supplemented by the initial values of $\overline{D}$, $\overline{\delta D^2}$ and $\overline{\delta D\delta\rho_s}$ at $t$=0.

Coupled equations (18) have the following physical interpretation (compare with modified approach [34]). The macroscopic state of the bulk sample with charged defects can be described by three parameters: $\overline{D}$, $\overline{\delta D^2}$ and $\overline{\delta D\delta\rho_s}$. The long-range order parameter $\overline{D}$ describes the ferroelectric ordering in the system, and the disorder parameter $\overline{\delta D^2}$ describes disordering caused by inner electric fields caused by charged non-homogeneities $\delta\rho_s$. The correlation $\overline{\delta D\delta\rho_s}$ determines the correlations between the movable screening cloud $\delta n$ and static charged defects $\delta\rho_s$.

As a resume to this section, we would like to stress that derived non-Hamiltonian system of coupled equations (18) is similar to the other well-known nonlinear systems of first order differential equations (e.g. the Lorenz system). Such dynamical systems possess chaotic regions, strange attractors as well as strongly non-ergodic behaviour and continuous relaxation time spectrum [24]. Any new system of such type demands a separate detailed mathematical study that was not the aim of this paper. Hereafter we discuss only the system behaviour near the equilibrium states far from the possible chaotic regions.

**5. Ferroelectric hysteresis.**

Let us consider the equilibrium solution of (18), which corresponds to the quasi-static external field changing. It is easy to check that the system (18) at $\Gamma=0$ is not the Hamiltonian one, i.e. it can not be directly obtained by varying of some free energy functional. This fact means that the obtained system of coupled equations is not equivalent to the equations of Landau-Khalatnikov type, which are derived from the minimization of Ginsburg-Landau potential.

In this section we demonstrate, how the ferroelectric hysteresis loop (i.e. the dependence of displacement $\overline{D}$ over the external field $E_0$) changes its shape under the presence of charged defects. First, let us rewrite equations (18) in dimensionless variables:

$$\frac{dD_m}{d\tau} - (1 - 3\Delta_D^2)D_m + D_m^3 = E_m \ ,$$

$$\frac{\tau_R}{2}\frac{d\Delta_D^2}{d\tau} - (\xi - 3D_m^2)\Delta_D^2 + \Delta_D^4 = -E_m K_{D\rho} + g\,R^2 \ , \qquad (19)$$

$$\tau_R\frac{dK_{D\rho}}{d\tau} - (\xi - 3D_m^2 - \Delta_D^2)K_{D\rho} = -E_m R^2 \ .$$

Here, $D_m = \overline{D}/D_S$, $\quad D_S = \sqrt{-\alpha/\beta}$, $\quad E_m = E_0/(-\alpha D_S)$, $\quad \Delta_D = \sqrt{\overline{\delta D^2}}/D_S$, $K_{D\rho} = \overline{\delta D\,\delta\rho_s}/\overline{\rho}_s D_S$, $\quad R^2 = \eta\,\overline{\delta\rho_s^2}/\overline{\rho}_s^2$, $\quad \tau = t/(-\alpha\Gamma)$, $\quad \tau_R = \Gamma/\Gamma_R$ and $g = -4\pi\,\overline{\rho}_s\,\kappa\,(1-\eta)\big/\eta\mu D_S^2$, $\xi = \alpha_R/\alpha$ at $\alpha < 0$. Hereafter we use the following initial conditions $D_m(\tau=0)=0$, $\Delta_D^2(\tau=0)=1$, $K_{D\rho}(\tau=0)=0$ and assume that $\Gamma \approx \Gamma_R$. We numerically analyze the solution of the system (19) in the following range of master parameters: $0<R<1$, $0\le g<1.5$, $-1\le\xi\le-1$.

The ferroelectric hysteresis loops $D_m(E_m)$ are represented in Figures 2-5 for the case of harmonic modulation of external field $E_m=E_{m,A}\sin(w\tau)$. Hereinafter we use the values of dimensionless frequency $w=-\alpha\Gamma\omega$ much less than unity. Selected value of the external field amplitude is the same order as the thermodynamic coercive field. The calculated hysteresis loops reveal the coercive field value well below the thermodynamic one. This result better agrees with experiments for bulk samples than predictions of classical Landau-Khalatnikov theory. Also the hysteresis loops shape strongly depends on $R$, $g$ and $\xi$ values, namely constricted, minor and double loops appear. The broadening, slight tilting and smearing of the loop shape under applied field frequency increasing can be explained by the dielectric losses increase (compare plots "a" and "b" in Figs.2-5).

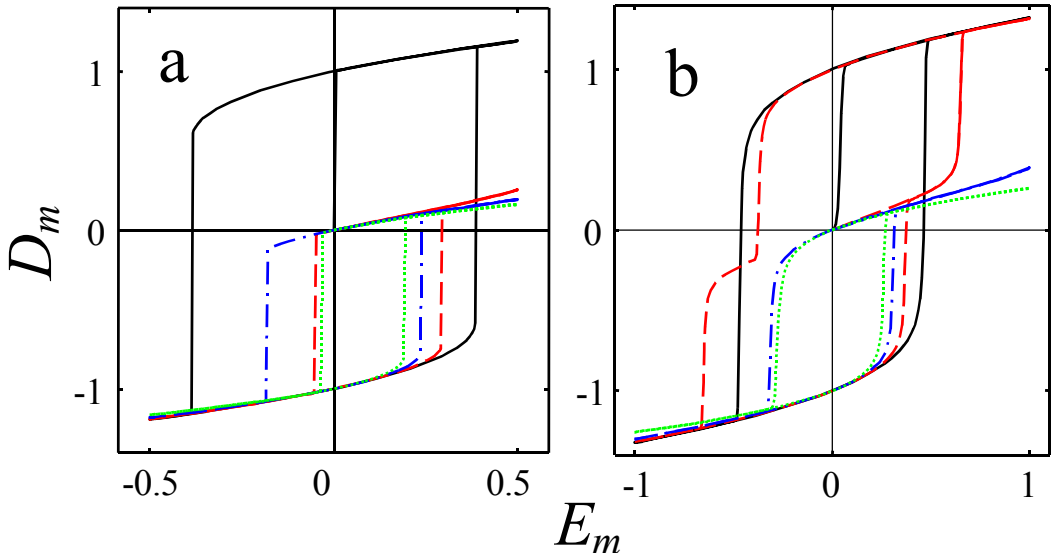

**Figure 2**. Hysteresis loops at frequencies $w=10^{-4}$ (plot a), $10^{-2}$ (plot b), $g\to0$, $\xi\to1$ and for different values of $R$: $R^2=0$ (solid curves), 0.1 (dashed curves), 0.4 (dot-dashed curves), 1 (dotted curves).

The typical hysteresis loops obtained at negligibly small diffusion coefficient (κ→0 and so $g$=0), correlation length (γ→0 and so ξ=1) and different $R$ values are shown in Fig. 2. It should be noted, that for $R$=1 and ξ→1, κ=0, $w$=0 hysteresis loops are absent [33]. For a given external field amplitude there is a critical value of parameter $R$. For the $R$ value above critical one instead of "full" loops we obtained only minor loops (see Fig. 2a). The realization of the "upper" or "lower" minor loop depends on the initial state of the sample. The similar minor loops were observed in PZT [1] and SBN: Ce single crystals [6]. Note, that minor loops exist only at ξ>0 and $g$=0. It is seen from Fig. 2b, that at higher frequencies minor loop obtained at small $R$ value transforms into the deformed "full" one.

The typical hysteresis loops obtained at non-zero diffusion coefficient (0<g<1), negative ξ value (-0.5<ξ<0) and different $R$ values are shown in Fig. 3. The loop becomes much slimmer and slightly lower under $R$ increasing (compare solid, dashed, dot-dashed and dotted curves).

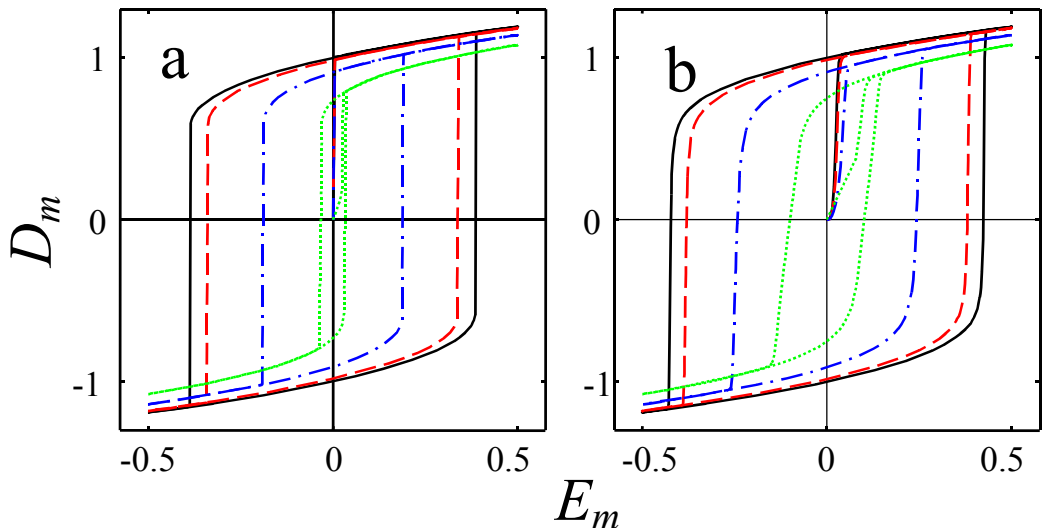

**Figure 3**. Hysteresis loops at low frequency $w$=10$^{-4}$ (plot a), 10$^{-2}$ (plot b), $g$=0.35, ξ=-0.5 and for different values of $R$: $R^2$=0 (solid curves), 0.1 (dashed curves), 0.5 (dot-dashed curves), 1 (dotted curves).

The Fig. 4 demonstrates the typical changes of hysteresis loop shape caused by increasing of charged defects density $\bar{\rho}_s$. The parameter $g \sim \bar{\rho}_s$ is always positive and typical ξ values are negative. The increase of $g$ value firstly leads to the significant coercive field decrease, then to the constriction appearance, subsequent transformation to the double loop and finally to the loop disappearance (compare different curves in Fig.4a). These changes of the loop shape smear under the frequency $w$ increasing (see Fig.4b).

The loop becomes much slimmer and slightly lower under $g$ increasing (compare solid, dashed, dot-dashed curves in Fig.4 with experimental data in Figs.3, 4 from [3] where Nd concentration increases in PZT film). The Figs.3, 4 demonstrate the drastic decrease of coercive field for g and $R$ values more than 0.5. This effect is somewhat similar to the known "square to slim transition" of the hysteresis loops in some relaxor ferroelectrics [35]. We can conclude that the increasing of charged defect concentration and its fluctuations lead to the coercive field value being significantly lower than thermodynamic coercive field [15].

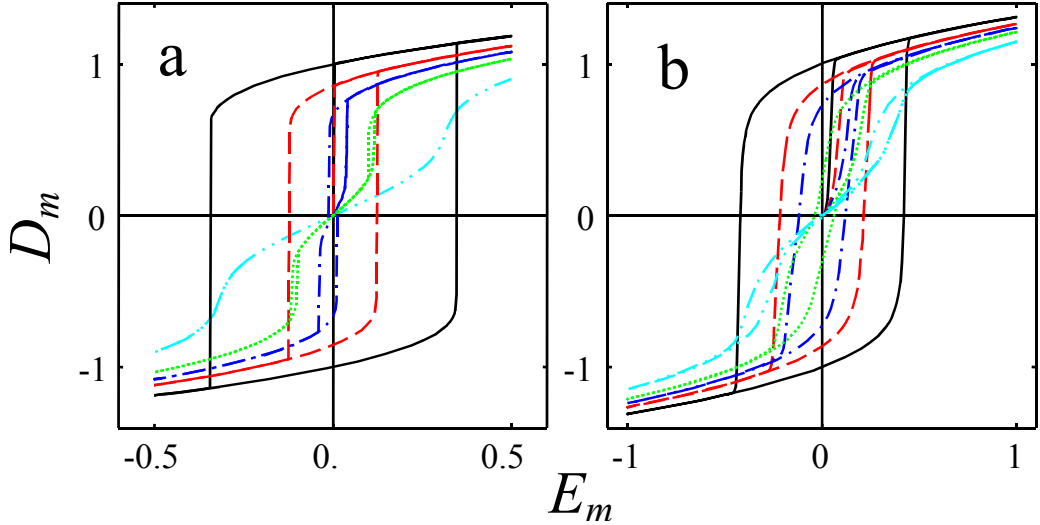

**Figure 4**. Hysteresis loops at low frequency $w$=10$^{-4}$ (plot a), 10$^{-2}$ (plot b), $R$ =0.5, $\xi$=-0.5 and for different values of $g$: $g$=0 (solid curve), 0.5 (dashed curves), 0.75 (dot-dashed curves), 1 (dotted curves), 1.5 (double dot-dashed curves).

The Fig. 5 demonstrates the typical changes of hysteresis loop shape caused by $\xi$ changing. The increase of $\xi$ value firstly leads to the coercive field decrease then to the constriction appearance, subsequent transformation to the double loop and finally to the loop disappearance (compare different curves in Fig.5a). These changes of the loop shape smear under the frequency $w$ increasing (see Fig.5b).

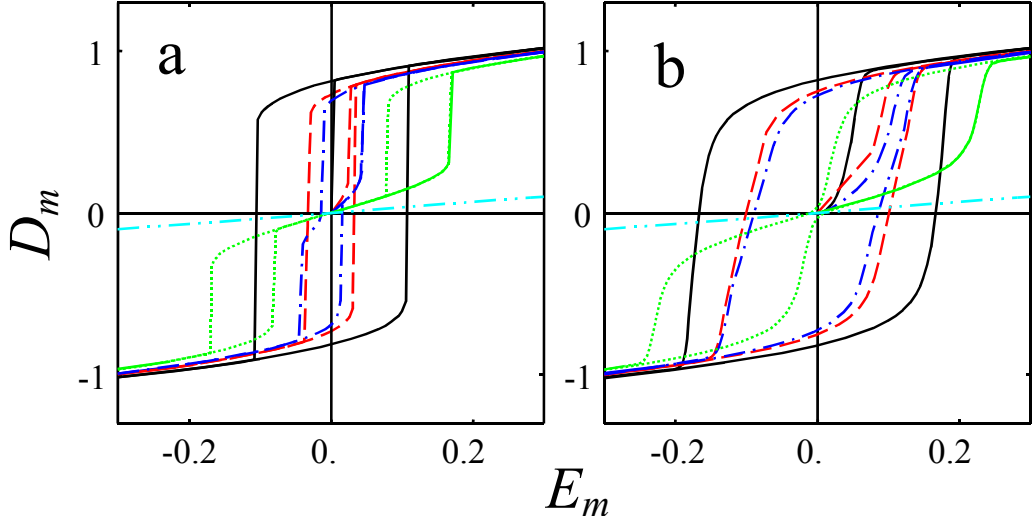

**Figure 5**. Hysteresis loops at frequency $w$=10$^{-4}$ (plot a), 10$^{-2}$ (plot b), $R$ =0.5, $g$ =0.7 and for different values of $\xi$=-1 (solid curves), -0.5 (dashed curves), -0.4 (dot-dashed curves), 0 (dotted curves), 1 (double dot-dashed curves).

It is seen from the Figs.4, 5 that $\xi$ and $g$ increase lead to the decrease of the sample coercive field and subsequent loop deformation from square loop to constricted loop, then to double one and loop vanishing. It should be noted, that the origin of the constricted or double loops in aged ferroelectric BaTiO$_3$ and (Pb,Ca)TiO$_3$ ceramics without charged impurities is caused by the mechanical clamping of spontaneous polarization switching [36] and so it lies outside our theoretical consideration. However, the constricted and double loops were observed in PLZT ceramics [3], [9], [10]. In this material La ions can be regarded as charged defects. In order to explain constricted and double loops evolution, we assume that baking at high pressures, annealing in special atmospheres or several hundreds of switching cycles can cause the significant changing of charged defects spatial distribution (e.g. the characteristic distance $d$ and value $R$). This effect leads to the changing of diffusion field amplitude $gR^2$ and $\xi$ value in accordance with (18) and (19) $\xi = 1 + \gamma / \alpha d^2$ . Thus within the framework of our model either appearance of constricted and double hysteresis loops in aged materials [9], [10] or their disappearance due to the "domain walls friction" [3] depends on the values of parameters $R$, $\xi$ and $g$ after sample treatment (compare dashed curve in Fig.1b, dot-dashed and dotted curves in Figs.4b,5b with Fig.1b in ref. [9] and Fig.4 in ref. [10]).

We can conclude that the increasing of charged defect concentration (as well as its fluctuations) leads to the decreasing of the sample coercive field, minor, constricted and double hysteresis loops related to the ferroelectric disordering. It seems that the method for determination of the parameters $R$, $\xi$ and $g$ (and then the material parameters of the sample) from the experimentally observed loop shape is desirable at least for the verification of the proposed phenomenological model.

### Dicussion

We have proposed the phenomenological description of polarization switching peculiarities in ferroelectric semiconductors with charged defects and prevailing extrinsic conductivity. Exactly we have modified Landau-Ginsburg approach for the aforementioned inhomogeneous systems and obtained the system of coupled equations (18). Both the experimentally observed coercive field values well below the thermodynamic one [15] and the various hysteresis loop deformations (minor, constricted and double loops [1], [2], [3], [6], [7], [9], [10]) have been obtained in the framework of our model.

The derived system of coupled equations (18) determines the evolution of order parameters. Mean displacement $\overline{D}$ is the order parameter. The quantitative degree of this disorder is the parameter $\sqrt{\overline{\delta D^2}}$ characterizing the inhomogeneity of the displacement distribution. It has been shown, that electric field fluctuations caused by charged defects lead to the ferroelectric disordering of the considered inhomogeneous system. Our calculations proved that the ferroelectric hysteresis loop disappears at the critical values of master parameters $R$, $g$, $\xi$. Probably this means that the phase transition into disordered state takes place below this threshold, e.g. the sample with defects splits into the polar regions (domains or Cross regions [35]).

Solving the system of coupled equations (18) one can get the information about system ordering as a whole, without defining space distribution of the appeared displacement inhomogeneities, domain walls characteristics, correlation radius of Cross regions or sizes of originated microdomains. In order to obtain this kind of information one has to solve the system of equations (9)-(12) with the specified distribution of defects, but the consideration of this problem was not the purpose of the present paper.

The size effects and the depolarization field influence [19], [20] have been neglected in the bulk of the sample due to enough sample thickness and free charge carriers influence. We suppose that inhomogeneous mechanical stresses arisen near defects are small enough or compensated by the sample treatment. Our theory qualitatively describes coercive field decreasing and various hysteresis loops deformation but not the aging process seen as the loop degradation. This may be related to the fact that possible evolution of the charge fluctuations caused by the relaxation/origin of internal stresses around defects was not taken into account in our model. These problems as well as the calculation of the system dielectric response are in progress now.

**Acknowledgments.**
The author is greatly indebted to Prof. N.V. Morozovsky for frutfull discussions of the model and useful remarks to the manuscript.

**Appendix A**

Let us estimate the dependence of parameters $\eta$, $R^2$ and $g$ on the charged defects distribution characteristics. We suppose that charged regions have the size $r_0$ and average distance between them is $d$. $d$ value can be expressed via defect concentration $n_d$ and lattice constant $a$ as $d = a/\sqrt[3]{n_d}$. If charged defects are clusterized real distance between them can be several times greater than $1/\sqrt[3]{n_d}$. We consider the case $d>>r_0$, i.e. the defects concentration do not exceed several percents (see Fig. 1). For the sake of simplicity, we approximate the defects charge densities by isotropic Gaussian distributions.

$$\rho_s(\mathbf{r}) = \sum_{i=1}^{N} \rho_{mi} \exp\left[-\frac{(\mathbf{r}-\mathbf{r}_i)^2}{r_0^2}\right]. \qquad (A.1a)$$

The charge density of carriers screening clouds originated around charge defects can be estimated using the Debye potential $\varphi(\mathbf{r}) \sim \dfrac{\exp\left(-|\mathbf{r}-\mathbf{r}_i|/R_D\right)}{|\mathbf{r}-\mathbf{r}_i|}$ [37] and Green function method, namely we obtain that Gaussian approximation is valid at small external field amplitudes:

$$n(\mathbf{r}) \approx \sum_{i=1}^{N} n_{mi} \exp\left[-\frac{(\mathbf{r}-\mathbf{r}_i)^2}{(r_0 + R_D)^2}\right]. \qquad (A.1b)$$

In the hypothetical case, when all impurity atoms are identical and regularly distributed one obtains that $\rho_{mi} \equiv \rho_m$, $n_{mi} \equiv n_m$, $|\mathbf{r}_{i+1} - \mathbf{r}_i| \equiv d$. In the case when $d >> (r_0 + R_D)$ we obtain the estimations:

$$\overline{\rho}_s = \frac{3}{d^3}\int_0^d r^2 dr\, \rho_m \exp\left(-\frac{r^2}{r_0^2}\right) \approx \frac{3\sqrt{\pi}}{4}\rho_m\left(\frac{r_0}{d}\right)^3, \qquad (A.2)$$

$$-\overline{\rho}_s = \overline{n} = \frac{3}{d^3}\int_0^d r^2 dr\, n_m \exp\left(-\frac{r^2}{(r_0+R_D)^2}\right) \approx \frac{3\sqrt{\pi}}{4}n_m\left(\frac{r_0+R_D}{d}\right)^3, \qquad (A.3)$$

$$\overline{\rho_s(\mathbf{r})n(\mathbf{r})} \approx \frac{3\sqrt{\pi}\,\rho_m n_m}{4\left[1 + r_0^2/(r_0+R_D)^2\right]^{3/2}}\left(\frac{r_0}{d}\right)^3, \qquad \overline{\rho_s^2(\mathbf{r})} \approx \frac{3\sqrt{\pi}\,\rho_m^2}{4}\left(\frac{r_0}{\sqrt{2}\,d}\right)^3. \qquad (A.4)$$

With the help of above-mentioned assumptions and calculations, it is easy to find the correlations:

$$\overline{\delta\rho_s\,\delta n} \equiv \overline{(\rho_s - \overline{\rho}_s)(n - \overline{n})} = \overline{\rho_s\,n} + \overline{\rho}_s^2, \qquad \overline{\delta\rho_s^2} \equiv \overline{(\rho_s - \overline{\rho}_s)^2} = \overline{\rho_s^2} - \overline{\rho}_s^2. \qquad (A.5)$$

Using the derived formulas (A.2)-(A.5) it is easy to obtain that

$$\overline{\delta\rho_s(\mathbf{r})\,\delta n(\mathbf{r})} \approx -\overline{\rho}_s^2\left(\frac{4(d/r_0)^3}{3\sqrt{\pi}\left[(1 + R_D/r_0)^2 + 1\right]^{3/2}} - 1\right), \qquad (A.6)$$

$$\overline{\delta\rho_s^2(\mathbf{r})} \approx \overline{\rho}_s^2\left(\frac{4}{3\sqrt{\pi}}\left(\frac{d}{\sqrt{2}\,r_0}\right)^3 - 1\right). \qquad (A.7)$$

Using the definitions (13) for $\eta$ ($\overline{\delta\rho_s\delta n} \approx -\eta\overline{\delta\rho_s^2}$) and (19) for $R^2$, g ($R^2 = \eta\overline{\delta\rho_s^2}/\overline{\rho}_s^2$, $g \sim \overline{\rho}_s(1-\eta)/\eta$)

$$\eta \approx \left(\frac{2}{\left(1+R_D/r_0\right)^2+1}\right)^{3/2}, \qquad R^2 \approx \frac{4}{3\sqrt{\pi}}\left(\frac{d}{r_0}\right)^3 \frac{1}{\left(\left(1+R_D/r_0\right)^2+1\right)^{3/2}}. \qquad (A.8)$$

Taking into account that $\overline{\rho}_s \sim n_d$, $d \sim 1/\sqrt[3]{n_d}$, $R_D \sim 1/\sqrt{n_d}$, one obtains that $R_D/r_0 = \sqrt{n_0/n_d}$ ($n_0$ is some characteristic constant concentration) and the following dependencies are valid

$$\eta \sim \frac{1}{\left(\left(1+\sqrt{n_0/n_d}\right)^2+1\right)^{3/2}}, \qquad R^2 \sim \frac{\eta}{n_d}, \qquad g \sim n_d(1-\eta)/\eta. \qquad (A.9)$$

However, when defects concentration becomes very small, e.g. $d{\to}\infty$ and $n_d \to 0$, intrinsic conductivity could not be neglected in comparison with extrinsic one and therefore the obtained expressions (A.8)-(A.9) become incorrect.

### Appendix B

The equations for $\overline{\delta D^2}$ and $\overline{\delta D\delta\rho_s}$ obtained directly from (12) have the form:

$$\frac{\Gamma}{2}\frac{\partial}{\partial t}\overline{\delta D^2} + \left(\alpha+3\beta\overline{D}^2(t)\right)\overline{\delta D^2} + \beta\overline{\delta D^4} = \gamma\overline{\delta D\frac{\partial^2\delta D}{\partial \mathbf{r}^2}} + \overline{\delta D\delta E_z}, \qquad (B.1)$$

$$\Gamma\frac{\partial}{\partial t}\overline{\delta D\delta\rho_s} + \left(\alpha+3\beta\overline{D}^2(t)\right)\overline{\delta D\delta\rho_s} + \beta\overline{\delta D^3\delta\rho_s} = \gamma\overline{\delta\rho_s\frac{\partial^2\delta D}{\partial \mathbf{r}^2}} + \overline{\delta\rho_s\delta E_z}. \qquad (B.2)$$

Under multiplying (12) over $\delta D^2$ or $\delta\rho_s\delta D$ and averaging allowing for (5) - (6) one obtains that

$$\overline{\delta D^4} = \left(\overline{\delta D^2}\right)^2, \qquad \overline{\delta D^3\delta\rho_s} = \overline{\delta D^2}\ \overline{\delta D\delta\rho_s}. \qquad (B.3)$$

Taking into account that the average period of the inhomogeneities distribution is $d$ (see Fig. 1), one obtains that $\partial/\partial\mathbf{r} \sim i/d$ and so

$$\gamma\overline{\delta D\frac{\partial^2\delta D}{\partial \mathbf{r}^2}} \sim -\frac{\gamma}{d^2}\overline{\delta D^2}, \qquad \gamma\overline{\delta\rho_s\frac{\partial^2\delta D}{\partial \mathbf{r}^2}} \sim -\frac{\gamma}{d^2}\overline{\delta\rho_s\delta D}. \qquad (B.4)$$

Let us express the field variation $\delta E_z$ via $\delta D$ and $\delta\rho_s$. In accordance with (17):

$$\delta E_z \approx \frac{\partial}{\partial t}\frac{\delta D}{4\pi\mu\overline{\rho}_s} - E_0(t)\left(\frac{\delta\rho_s}{\overline{\rho}_s} - \frac{div(\delta\mathbf{D})}{4\pi\overline{\rho}_s}\right) + \frac{\kappa}{\mu}\frac{\partial}{\partial z}\frac{\delta\rho_s}{\overline{\rho}_s} - \frac{\kappa}{\mu}\frac{\partial}{\partial z}\frac{div(\delta\mathbf{D})}{4\pi\overline{\rho}_s} + \frac{\delta n\,\delta E_z - \overline{\delta n\,\delta E_z}}{\overline{\rho}_s}$$
$$(B.5)$$

For a thick sample with equivalent boundaries $z = \pm\ell$ the term $\overline{\frac{\partial}{\partial z}\delta D^2}$ is equal to zero. In accordance with comments to (6) $\overline{\delta n\,\delta E_z^2} = 0$ and $\overline{\delta\rho_s\delta n\,\delta E_z} = 0$, thus one can derive from (B.5) the following approximations for the correlations:

$$\overline{\delta D\delta E_z} = \frac{1}{8\pi\mu\overline{\rho}_s}\frac{\partial}{\partial t}\overline{\delta D^2} - E_0(t)\frac{\overline{(\delta\rho_s\delta D)}}{\overline{\rho}_s} + \frac{\kappa}{\mu}\overline{\left(\delta D\frac{\partial}{\partial z}\frac{\delta\rho_s}{\overline{\rho}_s}\right)} - \frac{\kappa}{4\pi\overline{\rho}_s\mu}\overline{\delta D\frac{\partial}{\partial z}div(\delta\mathbf{D})}. \quad (B.6)$$

In accordance with (13) the term $\dfrac{\kappa}{\mu}\overline{\left(\delta D \dfrac{\partial}{\partial z}\dfrac{\delta \rho_s}{\overline{\rho}_s}\right)} = -\dfrac{\kappa}{\mu \overline{\rho}_s}\overline{\left(\delta \rho_s \dfrac{\partial}{\partial z}\delta D\right)}$ can be estimated as

$-\dfrac{4\pi\kappa}{\mu \overline{\rho}_s}\overline{\delta \rho_s(\delta \rho_s + \delta n)} \approx -(1-\eta)\dfrac{4\pi\kappa}{\mu \overline{\rho}_s}\overline{\delta \rho_s^2} = 16\pi^2 R_D^2(1-\eta)\overline{\delta \rho_s^2}$. Taking into account that the average period of the inhomogeneities distribution is $d$ (see Fig. 1), the last term in the right-hand side of (B.6) can be estimated as $\dfrac{\kappa}{4\pi\overline{\rho}_s\mu}\overline{\dfrac{\partial \delta D}{\partial z}div(\delta \mathbf{D})} \sim \dfrac{\kappa}{4\pi\overline{\rho}_s\mu}\overline{\left(\dfrac{\partial \delta D}{\partial z}\right)^2} = -\dfrac{R_D^2}{d^2}\overline{\delta D^2}$.

Thus the correlation

$$\overline{\delta D \delta E}_z = \dfrac{1}{8\pi\mu \overline{\rho}_s}\dfrac{\partial}{\partial t}\overline{\delta D^2} - E_0(t)\dfrac{\overline{(\delta \rho_s \delta D)}}{\overline{\rho}_s} - 16\pi^2 R_D^2(1-\eta)\overline{\delta \rho_s^2} - \dfrac{R_D^2}{d^2}\overline{\delta D^2}. \quad (B.7)$$

For a thick sample with equivalent boundaries $z = \pm \ell$ the term $\overline{\left(\dfrac{\partial}{\partial z}\dfrac{\delta \rho_s^2}{\overline{\rho}_s}\right)}$ is equal to zero and

therefore we obtain from (B.5) that

$$\overline{\delta \rho_s \delta E}_z = \dfrac{1}{4\pi\mu \overline{\rho}_s}\dfrac{\partial}{\partial t}\overline{\delta \rho_s \delta D} + E_0(t)\dfrac{\overline{\delta \rho_s \delta n}}{\overline{\rho}_s} - \dfrac{\kappa}{4\pi\overline{\rho}_s\mu}\overline{\delta \rho_s \dfrac{\partial}{\partial z}div(\delta \mathbf{D})}. \quad (B.8)$$

In accordance with (13) the last two terms in the right-hand side of (B.6) can be estimated as

$E_0(t)\dfrac{\overline{\delta \rho_s \delta n}}{\overline{\rho}_s} \approx -\eta\dfrac{\overline{\delta \rho_s^2}}{\overline{\rho}_s}E_0(t)$, $\dfrac{\kappa}{4\pi\overline{\rho}_s\mu}\overline{\delta \rho_s \dfrac{\partial}{\partial z}div(\delta \mathbf{D})} \sim -\dfrac{\kappa}{4\pi\overline{\rho}_s\mu d^2}\overline{\delta \rho_s \delta D} = \dfrac{R_D^2}{d^2}\overline{\delta \rho_s \delta D}$ and

so:

$$\overline{\delta \rho_s \delta E}_z = \dfrac{1}{4\pi\mu \overline{\rho}_s}\dfrac{\partial}{\partial t}\overline{\delta \rho_s \delta D} - \eta\dfrac{\overline{\delta \rho_s^2}}{\overline{\rho}_s}E_0(t) - \dfrac{R_D^2}{d^2}\overline{\delta \rho_s \delta D}. \quad (B.9)$$

Thus gradient terms in (B.1)-(B.2) can be either neglected at $\left(\gamma + R_D^2\right)/d^2 << \alpha$ or the coefficient $\alpha$ can be renormalized as $\alpha \to \alpha_R = \left(\alpha + \left(\gamma + R_D^2\right)/d^2\right)$. Using (B.3), (B.4), (B.7),

(B.9) we obtain the equations (18b) and (18c) from the equations (B.1) and (B.2).